\documentclass[aps, prb, superscriptaddress, twocolumn, amsmath,amssymb,notitlepage,nobalancelastpage,10pt,longbibliography]{revtex4-2}

\usepackage{amsmath}
\usepackage{verbatim}
\usepackage{graphicx}
\usepackage{color}
\usepackage{physics}
\usepackage{makecell}
\usepackage{yfonts}
\usepackage{soul}
\usepackage[normalem]{ulem}
\usepackage{dsfont}

\usepackage[colorlinks=true,citecolor=blue,linkcolor=blue,urlcolor=blue]{hyperref}
\usepackage{subfigure}

\usepackage{color}
\definecolor{darkgreen}{rgb}{0,0.60,.2}

\begin{document}

\title{Deconfined quantum critical points in fermionic systems with spin-charge separation}

\author{Niccol\`o Baldelli}
\affiliation{ICFO - Institut de Ci\`encies Fot\`oniques, The Barcelona Institute of Science and Technology, 08860 Castelldefels (Barcelona), Spain}
\author{Arianna Montorsi}
\affiliation{Institute for Condensed Matter Physics and Complex Systems,
DISAT, Politecnico di Torino, I-10129 Torino, Italy}
\author{Sergi Juli\`a-Farr\'e}
\affiliation{ICFO - Institut de Ci\`encies Fot\`oniques, The Barcelona Institute of Science and Technology, 08860 Castelldefels (Barcelona), Spain}
\author{Maciej Lewenstein}
\affiliation{ICFO - Institut de Ci\`encies Fot\`oniques, The Barcelona Institute of Science and Technology, 08860 Castelldefels (Barcelona), Spain}
\affiliation{ICREA, Pg. Llu\'is Companys 23, 08010 Barcelona, Spain}
\author{Matteo Rizzi}
\affiliation{Institute for Theoretical Physics, University of Cologne, D-50937 K\"oln, Germany}
\affiliation{Forschungszentrum J\"ulich GmbH, Institute of Quantum Control,
Peter Gr\"unberg Institut (PGI-8), 52425 J\"ulich, Germany}
\author{Luca Barbiero}
\affiliation{Institute for Condensed Matter Physics and Complex Systems,
DISAT, Politecnico di Torino, I-10129 Torino, Italy}

\begin{abstract}
Deconfined quantum critical points are intriguing transition points not predicted by the Landau–Ginzburg-Wilson symmetry-breaking paradigm which are usually identified by the appearance of a continuous phase transition between locally ordered phases. Here, we reveal the presence of deconfined quantum critical points with unexplored properties. Contrary to previously known examples, we show that the phenomenon of spin-charge separation peculiar to interacting low dimensional fermions can allow for the appearance of partially gapped deconfined quantum critical points. We first infer this point by performing a field theory analysis of generic one-dimensional fermionic systems in the low energy limit. Subsequently, we derive a microscopic model where phase transitions between different locally ordered phases can take place. Here, by performing a numerical analysis we explicitly derive, among others, the gaps, local order parameters and correlation functions behavior, supporting the presence of partially gapped deconfined quantum critical points. Our results thus provide new interesting insights on the widely investigated topic of quantum phase transitions.
\end{abstract}

\maketitle

\section{Introduction}
The mechanism by which one physical state is transformed
into a different one, namely a phase transition, represents one of the most fundamental concepts in several areas of physics ~\cite{Domb1972,Sachdev_2011}. In many cases, phase transitions are realized when an order parameter capturing the properties of a physical state abruptly vanishes, thus signaling the appearance of a phase with new features. This specific mechanism is usually defined as a first order or, equivalently, a discontinuous phase transition ~\cite{Binder1987}. Notably, a large variety of phase transitions displaying different behaviors exist. Paradigmatic examples are for instance the Berezinskii–Kosterlitz–Thouless ~\cite{Berezinsky1970,Kosterlitz_1973}, Ising ~\cite{Landau1937} and, the more exotic, symmetric mass generation ~\cite{sym14071475} phase transitions.\\Within this diversity, the Landau–Ginzburg-Wilson symmetry-breaking paradigm~\cite{Landau_ssb,wilson_ssb} rigorously predicts the appearance of discontinuous phase transitions connecting locally ordered (LO) phases with different symmetry properties. However, pioneering analysis proved~\cite{Senthil2004,Senthil2004v2} that it is possible to go beyond this paradigm. In particular, quantum fluctuations can generate second order continuous phase transitions between distinct gapped LO phases. This implies the presence of a gap closing, or equivalently, the simultaneous and continuous vanishing of the two order parameters characterizing the different LO phases, at a single point: the deconfined quantum critical point (DQCP)~\cite{senthil2023deconfined}. As proved for two-dimensional systems~\cite{Senthil2004,Senthil2004v2}, here very fascinating phenomena including charge fractionalization and emergent gauge fields can appear. In addition, DQCPs are further characterized by the phenomena of symmetry enrichment, meaning that they exhibit symmetry properties not present in any of the two LO phases they connect. This extraordinary variety of physical effects generated an intense theoretical effort which unveiled DQCPs in two-dimensional spin~\cite{Sandvik2007,Jiang_2008,Lou2009,Banerjee2010,Sandvik2010,Harada2013,Chen2013,Nahum2015,Shao2016,Wang2017a,Lee2019,Wang2022,song2023deconfined} and fermionic~\cite{Zi2019,Assaad2016,Liu2023,Yuan2023} models, as well as in 3D~\cite{Charrier2008,Sreejith2015, Mudry2019}, and 1D~\cite{Jiang2019,Roberts2019,Huang2019,Mudry2019,Roberts2021,lee2022,zhang2023,Baldelli2024,Romen2024} Hamiltonians. Notably, these previous studies all focused on systems where only a gap dictates the properties of phases with different symmetries. As a consequence, the gap closing at the transition point was always associated with fully gapless DQCPs, where all the correlations have algebraic decay. This is completely justified by the fact that, in the large majority of possible LO phases, the gaps capturing different degrees of freedom are always intrinsically related. Importantly, recent analysis \cite{Zhijin2022,Chester2024} based on conformal bootstrap \cite{Poland2019} challenged the existence of two dimensional DQCPs, while the same technique seems to confirm their presence in 3D \cite{chester2025}. These analysis thus motivate even more the characterization of DQCPs in 1D where highly accurate analytical and numerical techniques are available. Moreover, exact results have been recently produced \cite{zhang2023}. In such a context, fermionic systems deserve special attention. Indeed, many-body fermionic Hamiltonians can host the celebrated phenomenon of spin-charge separation ~\cite{Tomonaga1950,Luttinger1963,Haldane1981}. This accounts for the decoupling of spin and charge degrees of freedom which can therefore behave completely independently one from the other. Considering all these aspects, it becomes crucial to understand whether DQCPs can exist in the presence of spin-charge separation, and how this phenomenon may influence their properties.\\
\begin{figure}
    \centering
    \includegraphics[width=1.0\linewidth]{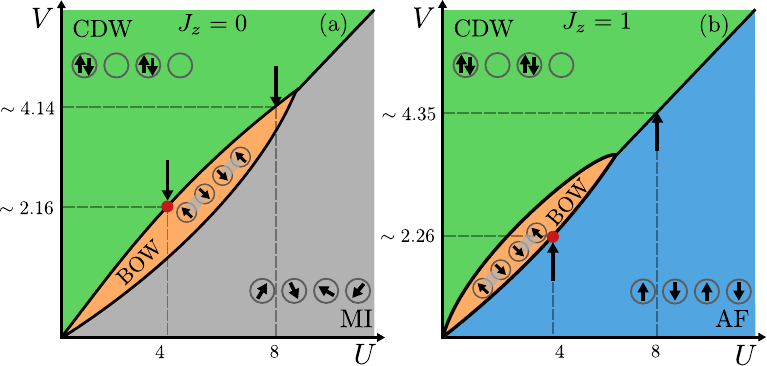}
    \caption{(a) Cartoon of the phase diagram of the microscopic Hamiltonian in Eq. \eqref{eq:ehm} for $J_z=0$. The phase diagram consists of a Mott insulator and bond-ordered-wave and a charge density wave. The arrows refer to the transition points studied in section \ref{bowcdw}. (b) Cartoon of the phase diagram of the microscopic Hamiltonian in Eq. \eqref{eq:ehm} for $J_z=1$. The phase diagram consists of an antiferromagnet and bond-ordered-wave and a charge density wave. The arrows refer to the transition points studied in sections \ref{afbow} and \ref{afcdw}. In both (a) and (b) the red dots indicate the presence of a partly gapped deconfined quantum critical point.}
    \label{phasediag}
\end{figure}
Here, we tackle this fundamental problem. Crucially, we find that spin-charge separation makes possible the appearance of a novel kind of DQCPs which, contrary to the previously known examples, have the special feature of being partially gapped. We prove this by considering in section \ref{sec:intro} the sine-Gordon model which well describes generic interacting one dimensional fermions in the low energy limit. By relying on a bosonization analysis ~\cite{Gogolin1998,Giamarchi2004}, we discuss how for conserved total particle number and spin magnetization, three possible fully gapped LO phases can take place: a charge density wave (CDW), an antiferromagnet (AF) and a bond-order-wave (BOW) insulator. 
Here, bosonization is able to prove the following key aspects:  \textit{a)} power law decay of specific correlation functions can occur at the phase transitions between the three different LO phases, thus proving the presence of at least one gap closing in one single point; \textit{b)} in two of these transition points, one gap closes while the other one remains finite as signaled by the long-range behavior of one nonlocal parity correlator; \textit{c)} at each of the three transition points one or two additional continuous symmetries emerge.\\In section \ref{micro:ham} we further enforce our findings by deriving a microscopic model where the three LO phases can take place. Here, in addition to bosonization, we perform a Variational Uniform Matrix Product States (VUMPS) ~\cite{Haegeman2013,Zauner2018} based numerical investigation. This method has the advantage that it can directly simulate systems in the thermodynamic limit, allowing for a systematic scaling procedure close to phase transitions \cite{Tagliacozzo2008,Rams2018,Vanhecke2019}. In this way, we unveil DQCPs where the concurrent disappearance of two local order parameters are related to the closing of a single gap, either the spin or charge one. Moreover, this analysis rules out a possible second order phase transition where both gaps vanish in one point. In this latter scenario, we indeed observe a first order phase transition as predicted by the Landau theory discuss the previous sentence. On one hand, this result proves the appearance of an emergent $U(1)\times U(1)$ symmetry in the channel associated with the gap closing. On the other, the presence of an additional finite gap induces long-range order of a parity correlator. Moreover, by performing a bond-dimension scaling extrapolation, we accurately extract the critical exponents which govern how the local order parameters vanish at the transition point. Thanks to this analysis, we find values not compatible with the Ising universality class thus supporting the appearance of one dimensional partially gapped DCQPs.\\On one hand, our results show that the celebrated phenomenon of spin-charge separation can largely enrich the physical properties of DQCPs and, on the other, our findings provide notable insights towards a deeper understanding of quantum phase transitions.

\begin{table*}
\setlength\extrarowheight{5pt}
\begin{tabular}{|c||c|c|c||c|c||c|c|c|}
\hline
& \multicolumn{3}{c||}{\textbf{Symmetry Protected Topological Phases}} & \multicolumn{2}{c||}{\textbf{Disordered Phases}} & \multicolumn{3}{c|}{\textbf{Locally Ordered Phases}}\\
\hline
& \makecell{Haldane \\ Insulator} & \makecell{Haldane \\ Liquid} & \makecell{Spin Haldane \\ Insulator} 
& \makecell{Luther-Emery\\ Liquid} & \makecell{Mott\\ Insulator} 
& \makecell{Charge Density \\ Wave} & Antiferromagnet & \makecell{Bond Order \\ Wave}\\ 
\hline
\makecell{Order \\ Parameter} 
& \makecell{\textit{nonlocal} \\ $\mathcal{O}_S^c$} 
& \makecell{\textit{nonlocal} \\ $\mathcal{O}_S^s$} 
& \makecell{\textit{nonlocal} \\ $\mathcal{O}_S^{c}$, $\mathcal{O}_S^{s}$} 
& \makecell{\textit{nonlocal} \\ $\mathcal{O}_P^s$} 
& \makecell{\textit{nonlocal} \\ $\mathcal{O}_P^c$} 
& \makecell{\textit{local} \\ $\mathcal{O}_{CDW}$} 
& \makecell{\textit{local} \\ $\mathcal{O}_{AF}$} 
& \makecell{\textit{local} \\ $\mathcal{O}_{BOW}$} \\
\hline
$\varphi_c$ 
& $\sqrt{\pi/8}$ 
& \textit{unpinned} 
& $\sqrt{\pi/8}$ 
& \textit{unpinned} 
& 0 
& $\sqrt{\pi/8}$ 
& 0 
& 0 \\
\hline
$\varphi_s$ 
& \textit{unpinned}
& $\sqrt{\pi/8}$
& $\sqrt{\pi/8}$
& 0 
& \textit{unpinned} 
& 0 
& $\sqrt{\pi/8}$
& 0 \\
\hline
$\Delta_c$ 
& $\neq0$ 
& $0$ 
& $\neq0$ 
& $0$ 
& $\neq0$ 
& $\neq0$ 
& $\neq0$ 
& $\neq0$  \\
\hline
$\Delta_s$ 
& $0$ 
& $\neq0$ 
& $\neq0$ 
& $\neq0$ 
& $0$ 
& $\neq0$ 
& $\neq0$ 
& $\neq0$ \\
\hline
\end{tabular}
\label{tab:table_phases}
\caption{List of all the possible gapped phases in one-dimensional fermionic systems with $U(1)$ charge and spin symmetry described by the Hamiltonian in Eq.\eqref{SG} and derived by studying the microscopic model in Eq.\eqref{eq:ehm}. These phases are distinguished by considering the presence of different gaps and their order parameters.}
\end{table*}

\section{Field Theory of interacting fermions}\label{sec:intro}
\medskip
\subsection{Derivation of phases with nonlocal or local order}

Interacting fermionic systems allow for a field theory description, also called bosonization ~\cite{Gogolin1998,Giamarchi2004}, accurate in the low energy limit. In particular, a generic lattice microscopic Hamiltonian conserving the total magnetization and particles number can be rewritten in the continuum limit as
two decoupled sine-Gordon models  ${\cal H}=\sum_{\nu=c,s}{\cal H}_\nu^{SG}$ capturing the spin ($s$) and charge ($c$) degrees of freedom 
\begin{equation}
\begin{split}
 {\cal H}^{SG}_\nu = \frac{1}{2} \int dx   \bigg[ v_\nu K_\nu (\nabla \vartheta_\nu(x))^2  +\frac{v_\nu}{K_\nu}  (\nabla \varphi_\nu(x))^2\\
 +\frac{g_\nu}{\pi^2 a^2} \cos(\sqrt{8\pi} \varphi_\nu(x))\bigg],
 \quad 
 \end{split}
 \label{SG}
 \end{equation}
see Appendix \ref{app:bosonization} for more details. Here, $v_\nu$, $K_\nu$ and $g_\nu$ are the excitation velocities, Luttinger parameters, and coupling amplitudes respectively, which depend on the microscopic Hamiltonian parameters. At the same time, the behavior of the bosonic fields $\varphi_\nu(x)$ and $\vartheta_\nu(x)$ governs the appearance of gapped phases, see \ref{app:bosonization} . Specifically, a finite gap $\Delta_\nu$ in the $\nu$ channel develops when $\varphi_\nu(x)$ is pinned to a specific value. Notice that in such a case the effective symmetry relative to the $\nu$ channel is $U(1)$ (see Appendix). On the other hand, if $\varphi_\nu(x)$ remains unpinned, the $\nu$ channel is gapless and an emergent $U(1)\times U(1)$ ~\footnote{notice that this may be enhanced to a full $SU(2)$ symmetry in case of symmetric choices of the microscopic parameters of a Hamiltonian\cite{Mudry2019}} symmetry appears. In this regard, eq. \eqref{SG} provides different solutions describing gapped phases.\\
 We start by discussing the following cases: \textit{a)} $\varphi_c=\sqrt{\frac{\pi}{8}}$ and $\varphi_s$ is unpinned; \textit{b)} $\varphi_s=\sqrt{\frac{\pi}{8}}$ and $\varphi_c$ is unpinned \footnote{Importantly, this scenario has also been described as a gapless SPT regime \cite{Keselman2015, Rakovszky2020}}; \textit{c)} $\varphi_c=\sqrt{\frac{\pi}{8}}$ and $\varphi_s=\sqrt{\frac{\pi}{8}}$. It turns out that gapped phases captured by such specific value $\sqrt{\frac{\pi}{8}}$ of the pinned field $\varphi_\nu(x)$ can be uniquely identified through the nonlocal string operator
\begin{eqnarray}
 O_S^{\nu}(j)={\rm e}^{i\pi \sum_{i<j}S_i^\nu} S_j^\nu\rightarrow O_S^{\nu}(x)\sim\sin(\sqrt{2\pi}\varphi_\nu(x)) \label{SOP}
\end{eqnarray}
in the microscopic and bosonized version respectively \cite{Montorsi2012,Barbiero2013}, where 

\begin{eqnarray}
S_j^s=n_{j\uparrow}-n_{j\downarrow},\quad S_j^c=1-n_{j}, 
\label{eq:spin_op} 
\end{eqnarray}
 are fermionic number operators. Here,  $n_j=n_{j,\uparrow}+n_{j,\downarrow}$ is the total number of particles. Specifically, states of matter which ordering is captured solely by a finite string order parameter
\begin{eqnarray}
 \mathcal{O}_S^{\nu}=\frac{1}{L}\sum_j\langle O_S^{\nu}(j) \rangle\rightarrow \mathcal{O}_S^{\nu}\sim\frac{1}{L}\int dxO_S^{\nu}(x) \label{SOPA}
\end{eqnarray}
can be recognized ~\cite{Montorsi2017} as symmetry protected topological (SPT) phases ~\cite{Pollmann2012,Tang2012,Wen2014,Gu2014,Senthil2015,Shapourian2017}. Based on the formal analogy with the SPT phases occurring in spin-$1$ Heisenberg ~\cite{haldane1983,affleck1987,barbiero2017} and Bose-Hubbard models ~\cite{DallaTorre2006,Rossini2012, Sugimoto2019,Fraxanet2022}, the \textit{a)} solution thus describes a charge gapped Haldane insulator. A spin gapped SPT regime also known as Haldane liquid ~\cite{Fazzini2019,Montorsi2020} emerges when only $\varphi_s=\sqrt{\frac{\pi}{8}}$ and $\varphi_c$ keeps fluctuating, i. e. \textit{b)}. Finally, \textit{c)} implies that both $\mathcal{O}_S^{\nu}\neq 0$ and therefore a fully gapped spin Haldane insulator phase analogous to the topological regime of spinful SSH models ~\cite{Anfuso2007,Manmana2012,Barbiero2018,julia-farre2022} takes place. Notably, this intrinsic nonlocality makes SPT phases the paradigmatic example of states escaping the Landau's theory of LO states of matter.\\Two additional phases characterized by the absence of local order, and therefore also not captured by the Landau's theory, occur when $\varphi_{c(s)}=0$ and $\varphi_{s(c)}$ is unpinned. Upon defining the nonlocal parity, also called in higher dimension disorder ~\cite{Kadanoff1971,Nussinov2009,Fradkin2017}, operators
\begin{eqnarray}
   O_P^{\nu}(j)= {\rm e}^{i\pi \sum_{i<j}S_i^\nu} \rightarrow O_P^{\nu}(x)\sim\cos(\sqrt{2\pi}\varphi_\nu(x)),  \label{POP}  
\end{eqnarray}
one can show that their expectation value 
\begin{eqnarray}
 \mathcal{O}^{\nu}_P=\frac{1}{L}\sum_j\langle O_P^{\nu}(j) \rangle\rightarrow \mathcal{O}^{\nu}_P\sim\frac{1}{L}\int dxO_P^{\nu}(x) \label{POPA}
\end{eqnarray}
represents the order parameters of such phases.
More in detail, these two solutions describe: a spin gapped Luther-Emery liquid (LEL) phase where $\varphi_s=0$ and $\varphi_c$ is unpinned so that only $\mathcal{O}_P^{s}\neq 0$; a Mott insulator (MI) where the charge and spin gaps are $\Delta_c\neq 0$, $\Delta_s=0$, thus reflecting the unpinning of $\varphi_s$ and the pinning of $\varphi_c$ to zero so that only $\mathcal{O}_P^{c}\neq 0$. Because of absence of any local order, we name both LE and MI as disordered phases.\\
 \begin{figure*}
    \centering
    \includegraphics[width=\textwidth]{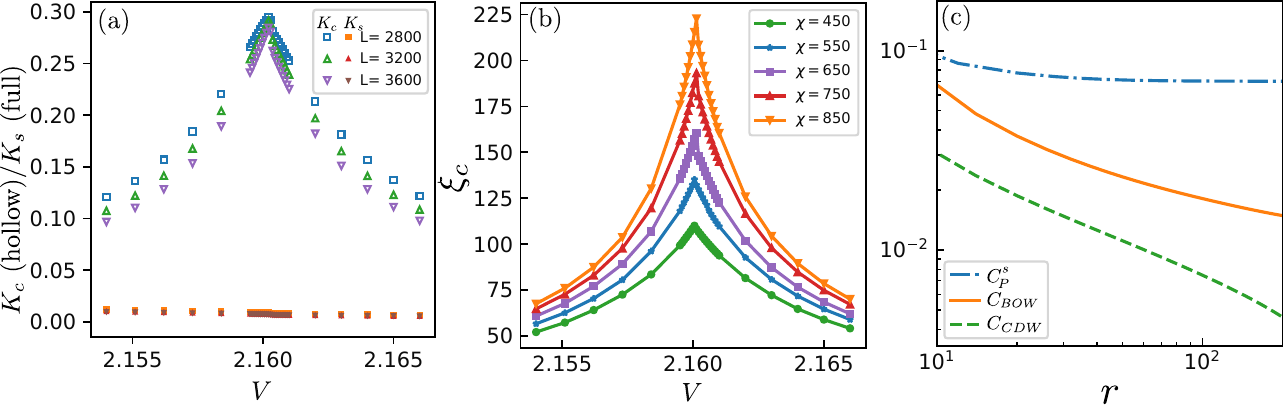}
    \caption{Study of the critical point between the BOW and the CDW phases appearing in Eq. \eqref{eq:ehm} for $U = 4$, $J_z=0$ and $t=1$. (a) Luttinger parameters $K_s$ and $K_c$. While the first is almost vanishing for all the considered values of $V$ thus proving that the spin gap remains always finite, $K_c$ shows a size independent cusp in one single point thus proving the charge gap closing. Results expressed as a function of a finite MPS of size $L$, constructed out of $L/2$ unit cells of the infinite MPS. (b) The correlation length $\xi_c$ shows a bond-dimension $\chi$-divergence in the form of a cusp only in one point, also signaling a closing of a gap. (c) Algebraic decay of $C_{CDW}$ and $C_{BOW}$ associated to the long-range order of the spin parity operator correlation $C_P^s$ at the CDW-BOW transition point. All the results have been obtained through VUMPS simulation by keeping a unit cell of two sites.}
    \label{fig:CDW_BOW}
\end{figure*}
Up to now, the field theory treatment characterized quantum regimes with no local orders. Notably, throughout the above bosonization analysis three fully gapped LO phases can be identified, see Fig. \eqref{phasediag} for a cartoon of this regimes. The first one turns out to be a charge density wave (CDW) corresponding to the solution $\varphi_c=\sqrt{\frac{\pi}{8}}$ and $\varphi_s=0$. Here, the relative local order parameter in his microscopic and bosonized version is represented by a finite expectation value of $S_j^c$, i.e.
\begin{eqnarray}
    \mathcal{O}_{CDW}&=&\frac{1}{L}\sum_j(-1)^j \langle S^c_j\rangle\rightarrow\\
    \mathcal{O}_{CDW}&\sim&\frac{1}{L}\int dx \langle\sin (\sqrt{2\pi} \varphi_c(x)) \cos(\sqrt{2\pi} \varphi_s(x))\rangle\nonumber .
    \label{eq:CDWOP}
\end{eqnarray}
This latter shows the appearance of a LO phase where pairs and empty sites are perfectly alternated thus proving the breaking of the discrete translation symmetry $c_{j\sigma}\rightarrow c_{j+1\sigma}$. The second LO regime refers to the solution $\varphi_c=0$ and $\varphi_s=\sqrt{\frac{\pi}{8}}$. This regime, also known as antiferromagnet (AF), appears when the discrete lattice spin flip invariance $c_{j,\sigma}\rightarrow c_{j,-\sigma}$ is broken, reflecting the perfect alternation between fermions with antiparallel spins. In this regard, the local order parameter capturing the AF phase is the expectation value of the staggered magnetization $S_j^s$   
\begin{eqnarray}
\mathcal{O}_{AF}&=&\frac{1}{L}\sum_j(-1)^j \langle S^s_j\rangle\rightarrow\\
\mathcal{O}_{AF}&\sim&\frac{1}{L}\int dx \langle\cos (\sqrt{2\pi} \varphi_c(x))\, \sin (\sqrt{2\pi} \varphi_s(x))\rangle\nonumber
\label{AFOP}
\end{eqnarray}
in its microscopic and bosonized version, respectively. Finally, bosonization predicts an additional LO phase when $\varphi_c=0$ and $\varphi_s=0$. This latter, usually called bond-order-wave (BOW) phase, appears when the the discrete site inversion symmetry is broken. As a consequence, an effective lattice dimerization is captured by the local order parameter
\begin{eqnarray}
\mathcal{O}_{BOW}&=&\frac{1}{L}\sum_{j}(-1)^{j}\langle B_j\rangle\rightarrow\\
\mathcal{O}_{BOW}&\sim&\frac{1}{L}\int dx \langle\cos (\sqrt{2\pi} \varphi_c(x))\, \cos (\sqrt{2\pi} \varphi_s(x))\rangle\nonumber
\label{eq:BOWOP}
\end{eqnarray}
with $B_j=\sum_\sigma(c^\dagger_{j,\sigma}c_{j+1,\sigma}+h.c.)$.

The complete classification of gapped and partly gapped many-body phases is reported in Table \ref{tab:table_phases}
\subsection{Beyond Landau phase transitions}
The previous analysis allowed us to define three LO regimes. It appears thus crucial to understand whether such phases can be connected through continuous or first order phase transitions. As already pointed out, in the former case the gap vanishes uniquely in one specific point and therefore algebraic decay of the two correlation functions having long-range or exponential decay in the two LO regimes should be predicted. 
In addition, one should distinguish the case where at these transitions only one of the two gaps closes and reopens from the case where both gaps vanish independently. 
While this latter scenario might occur at the AF-CDW transition as signaled by the different value of both $\varphi_s$ and $\varphi_c$ in AF and CDW, the constant value of $\varphi_s(\varphi_c)$ in BOW and CDW (AF) is compatible with the fact that only $\Delta_c(\Delta_s)=0$ at the BOW-CDW (AF) transition point. As a consequence, in this last case the presence of a second order transition will be further enriched by one gap $\Delta_\nu\neq 0$ and the corresponding nonlocal order parameter $\mathcal{O}_P^{\nu}$ will remain finite. Therefore, one might detect a LEL (MI) with spin (charge) parity order at the BOW-CDW (AF). Whereas, in presence of second order AF-CDW transition both gaps have to close and a Luttinger liquid (LL) behavior can take place. Only in this latter case all correlation functions would then display algebraic decay.\\
In order to better exploit this crucial point, it is convenient to first relate the decay of correlation functions at the transition points to the Luttinger parameters $K_\nu$. In this regards, see \ref{app:bosonization} for more details, bosonization shows $K_\nu\approx 0$ in presence of $\Delta_\nu\neq 0$ while $K_\nu$ remains thermodynamically finite in presence of a vanishing gap in the $\nu$ channel. This point is quite important as the Tomonaga Luttinger theory \cite{Tomonaga1950,Luttinger1963} predicts the $K_\nu$'s to induce in the MI, LEL, and LL phases different algebraic decay of the correlation functions $C_{X}(r)$ (with $X=CDW, AF, BOW$) having long-range order in the $X$ LO phase. Notably, these latter can have the same decay only in a fully gapless Luttinger liquid
\begin{equation}
    C_{AF}(r)\sim C_{BOW}(r)\sim C_{CDW}(r) \sim r^{-(K_c+K_s)}.
    \label{cLL} 
\end{equation}
As already discussed, this is the scenario potentially characterizing the AF-CDW phase transition. As a consequence, the expression in Eq. \eqref{cLL} suggests the continuous nature of this phase transition which therefore would be not captured by the Landau's theory. Moreover, bosonization finds that at the BOW-CDW transition solely $K_s=0$ and therefore the relation showing the presence of algebraic decay
\begin{equation}
C_{CDW}(r) \sim r^{-K_c}\sim C_{BOW}(r)
\label{Corcdwbow}
\end{equation}
holds. In addition, the fixed value $\varphi_s=0$ suggests that long-range order of the spin parity correlation $C_P^s(r)=\langle O^{s\dagger}_P(x)O_P^s(x+r)\rangle$ has to occur. Analogously, at the BOW-AF transition $K_c=0$, therefore
\begin{equation}
 C_{AF}(r) \sim r^{-K_s}\sim C_{BOW}(r)
 \label{corafbow}
\end{equation}
and, in addition, $\varphi_c=0$ so that $C_P^{c}(r)=\langle O^{c\dagger}_P(x)O_P^c(x+r)\rangle$ is long-range ordered. The above results clearly  point in the direction that exotic partially gapped DQCPs characterized by long-range order of nonlocal order parameter can exist at the BOW-CDW and BOW-AF transition while standard fully gapless DQCPs might appear at the CDW-AF critical point. 
\section{Microscopic Hamiltonian}
\label{micro:ham}
\noindent As the field theory analysis show evidences of possible very peculiar beyond-Landau phase transitions occurring in strongly interacting fermionic systems, we turn to confirming this intuition at a microscopical level. To do that, we design a lattice Hamiltonian where both the three LO regimes and phase transition between them can occur, see Fig. \eqref{phasediag}. In particular, while broken translational symmetry is usually induced by the intersite density-density interaction
\begin{equation}
H_{IS}(j)=S_{j}^cS_{j+1}^c,
\end{equation}
antiferromagnetic order can emerge in presence of the magnetic coupling 
\begin{equation}
H_{AF}(j)=S^s_jS_{j+1}^s.
\end{equation}
Here, the operators $S_{j}^c$ and $S^s_j$ are the ones defined in Eq.\eqref{eq:spin_op}. In addition, the competition between intersite and onsite interaction of the form of 
\begin{equation}
H_{OS}(j)=n_{j\uparrow}n_{j,\downarrow}
\end{equation}
can give rise to BOW ordering \cite{Nakamura2000}. Finally, away from strongly interacting perturbative regimes, quantum fluctuations are induced by tunneling process 
\begin{equation}
H_{TP}(j)=\sum_\sigma(c^\dagger_{j,\sigma}c_{j+1,\sigma}+h.c.).
\end{equation}
All these contributes and their relative strength are captured by the following Hamiltonian
\begin{equation}
 H=\sum_{j}[-tH_{TP}(j)+VH_{IS}(j)+J_{z}H_{AF}(j)+UH_{OS}(j)].
\label{eq:ehm}
\end{equation}
 Notably, this latter presents strong connections with the Hamiltonians \cite{julia-farre2022,guardado-sanchez_quench_2021,Su2023,carroll2024} under investigations in the context of ultracold atomic systems. As these setups are characterized by an unprecedented level of accuracy and versatility, here we restrict our analysis to the regime of half-filling and vanishing total magnetization, i.e.  $\sum_jS^{\nu}_j=0$ for both $\nu=c,s$.
\subsection{Bosonization analysis}\label{boso}
We can extract important information about possible beyond-Landau phase transitions, by applying the bosonization approach to this microscopic model. As shown in Appendix \ref{app:bosonization}, 
%
%
%
bosonization predicts that the transitions between different LO phases occur independently in the two channels 
but they are captured by the same line  
\begin{equation}
U+2J_z=2V \quad . \label{Jztrans}
\end{equation} 
Thus, along such line, a continuous CDW-AF transition is predicted, in principle compatible with a fully gapless DQCP. However, it is a well known fact that when the above analysis is complemented by other results, for instance by level crossing analysis \cite{Nakamura2000}, the actual transition lines observed in the two channels split: the BOW-CDW and BOW-AF transitions will occur in the weak coupling region, while they will merge into the CDW-AF transition at intermediate couplings only. If this is the case, in the weakly coupled regime this theory turns out to be compatible with the presence of the partially gapped DQCPs envisaged previously. 
\\At the transition lines Eq. \eqref{Jztrans}, one can show \ref{app:bosonization} that $K_c\approx 1-\frac{2V}{\pi t}$ and $K_s\approx1-\frac{2 J_z}{\pi t}$, differing from each other for any non-vanishing $V\neq J_z$. These values for the Luttinger parameters will be similar at the CDW-AF, or BOW-AF and BOW-CDW transitions. On the other hand, depending on the actual gap closing, eqs. \eqref{cLL}, \eqref{Corcdwbow}, and \eqref{corafbow} predict a completely different dependence on $K_\nu$ in the power law decay of correlation functions. Thus we expect that the latter will potentially be valuable indicators of the possible type of DQCP. 
\begin{figure}
    \centering
    \includegraphics[width=0.75\linewidth]{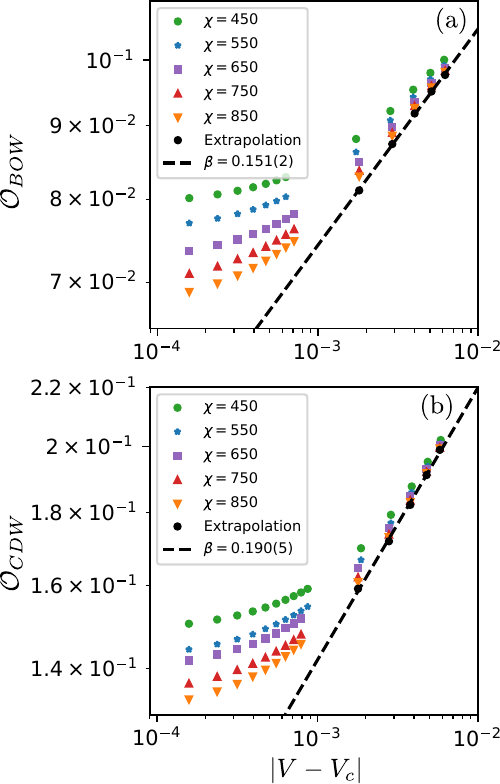}
    \caption{Scaling of the order parameters around the BOW-CDW transition relative to the model in Eq. \eqref{eq:ehm} for $U = 4$, $J_z=0$ and $t=1$ as a function of the bond dimension $\chi$. Both the (a) BOW and the (b) CDW order parameters vanish continuously at transition. By extrapolating the limit $\chi\rightarrow\infty$, it is possible to extract the critical exponents $\beta_{CDW},\beta_{BOW}\neq 1/8$, thus excluding an Ising transition. The details of the extrapolation are detailed in Appendix \ref{app: extrap}}
    \label{fig:CDW_BOW_order}
\end{figure}
\subsection{VUMPS analysis}
Since bosonization is believed to provide accurate results in the low energy limit, in the following we employ numerical calculations to explore the possible appearance of DQCPs away from this regime. To unveil this point, we perform a numerical analysis based on the VUMPS technique. Here, the state is described by an infinitely repeating unit cell of tensors defined on the sites of the lattice. The tensors are then variationally optimized with a procedure similar to DMRG \cite{White1992}, by using the rest of the (infinite) system as an environment. Compared to infinite DMRG, VUMPS has the advantage to completely replace the full environment at each step, making its convergence significantly faster. This is crucial especially at critical points, where a larger bond dimension is required to get accurate results. Moreover, the absence of boundaries in the variational state makes it easier to escape metastable states that can become more favourable around the critical points in study, where two states that break different symmetries have comparable energies. Indeed, the use of VUMPS has proved critical in the analysis of DQCPs in other systems \cite{Roberts2019, Baldelli2024}. 

\subsubsection{BOW-CDW phase transition}\label{bowcdw}
As shown in Fig. \eqref{phasediag}(a), for $J_z=0$, the Hamiltonian in Eq.\eqref{eq:ehm} is the known extended Fermi-Hubbard model \cite{Nakamura2000,Sengupta2002,Sandvik2004,Ejima2007,Zhang2004}. For dominating $U/t$, the system is in a MI phase. 
\begin{figure}
    \centering
    \includegraphics[width=0.7\linewidth]{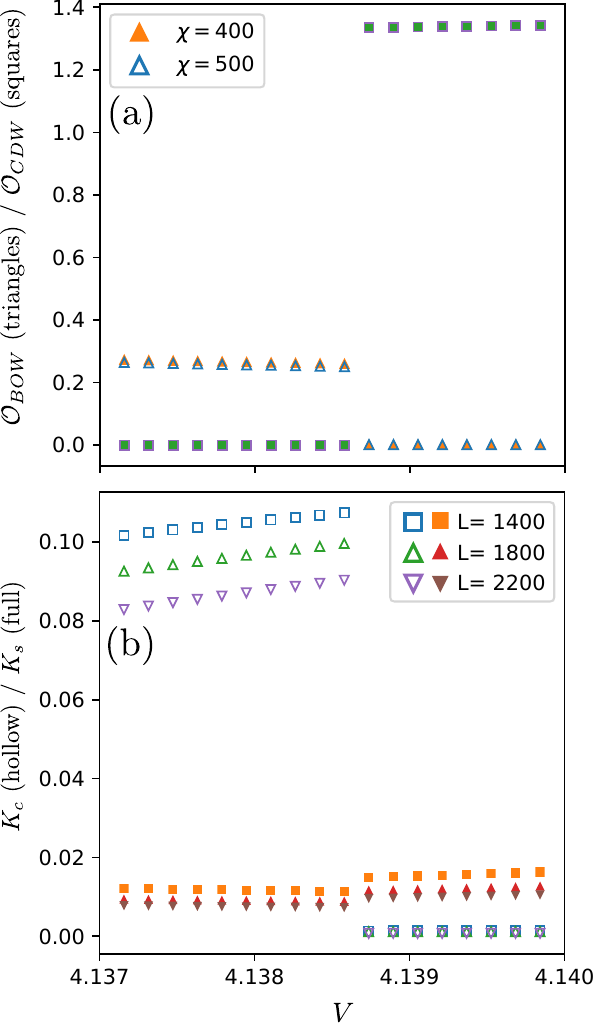}
    \caption{First order phase transition between the BOW and the CDW phases in the model Eq. \eqref{eq:ehm} for $U=8, J_z=0$ and $t=1$. (a) Values of the order parameters $\mathcal{O}_{BOW}$ and $\mathcal{O}_{CDW}$ for different dimension $\chi$; (a) Values of the Luttinger parameter $K_c$ and $K_s$ at fixed bond dimension. The size $L$ refers to a finite MPS constructed out of $L/2$ unit cells.}
    \label{fig:CDW_BOW_firstorder}
\end{figure}
The situation changes drastically for larger values of $V$. Here, two LO phases can take place: a BOW phase when $V\sim U$ are away from strongly interacting perturbative regimes, and a CDW for dominant intersite repulsion. 
\begin{figure*}
    \centering
    \includegraphics[width=\textwidth]{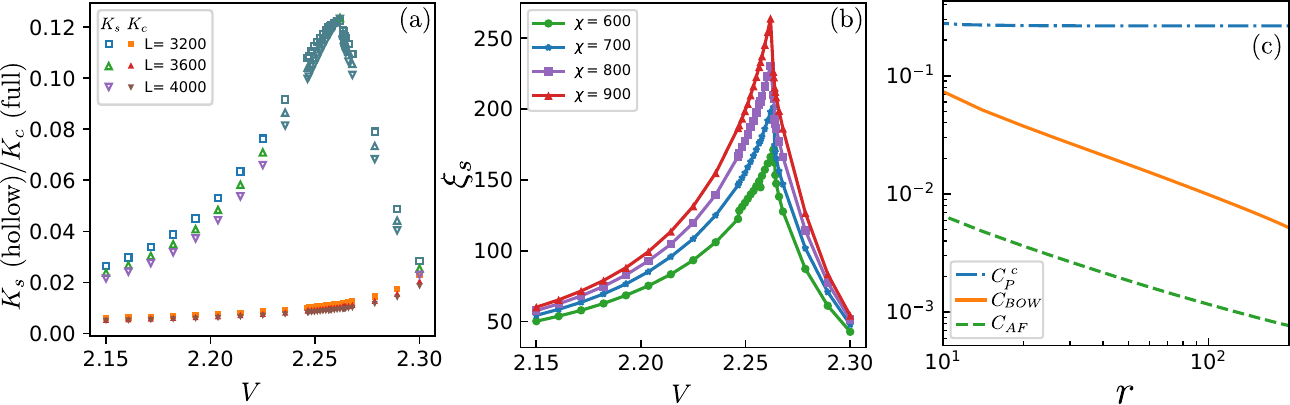}
    \caption{Study of the critical point between the BOW and the AF phases appearing in Eq. \eqref{eq:ehm} for $U =4$, $J_z=t=1$. (a) Luttinger parameters $K_c$ and $K_s$. While the first is almost vanishing for all the considered values of $V$ thus proving that the spin gap remains always finite, $K_s$ shows a size independent cusp in one single point thus proving the spin gap closing. Results expressed as a function of a finite MPS of size $L$, constructed out of $L/2$ unit cells of the infinite MPS. (b) The correlation length $\xi_s$ shows a bond-dimension $\chi$-divergence in the form of a cusp only in one point, also signaling a closing of a gap. (c) Algebraic decay of $C_{AF}$ and $C_{BOW}$ associated to long-range order of the charge parity correlator $C_P^c$ at the AF-BOW transition point. All the results have been obtained through VUMPS simulation by keeping a unit cell of two sites.}
    \label{fig:AF-BOW_crit}
\end{figure*}
To certify that at the BOW-CDW critical point a gap closing occurs and that it only happens in the charge sector, we extrapolate the Luttinger parameters $K_\nu$. Specifically, we extract their thermodynamic value by computing the static structure factors $P_\nu(q) = \frac{1}{L} \sum_{r} e^{iqr} \langle S_j^\nu S_{j+r}^\nu \rangle$ which relate to the Luttinger constants through the equation
\begin{equation}
    K_\nu = \lim_{q\rightarrow 0} \frac{P_\nu(q)}{q} \sim 2\pi\lim_{L\rightarrow\infty} \frac{P_\nu(q=2\pi/L)-P_\nu(q=0)}{L}. 
\end{equation}
As discussed in the previous sections, this quantity is expected to be zero for a finite  $\Delta_\nu$, and finite in the case of a gapless $\nu$-channel. To perform the finite-size scaling, we construct multiple finite MPS of length $L$ by repeating $L/2$ unit cells obtained from the VUMPS optimization.
Although our analysis considers relatively large interactions $U/t=4$, our VUMPS simulations provide results compatible with the field theory. On one hand, in Fig.\ref{fig:CDW_BOW}(a) we indeed find that $K_s\approx 0$ at the CDW-BOW transition point, meaning that $\Delta_s\neq0$.
\begin{figure}
    \centering
    \includegraphics[width=0.75\linewidth]{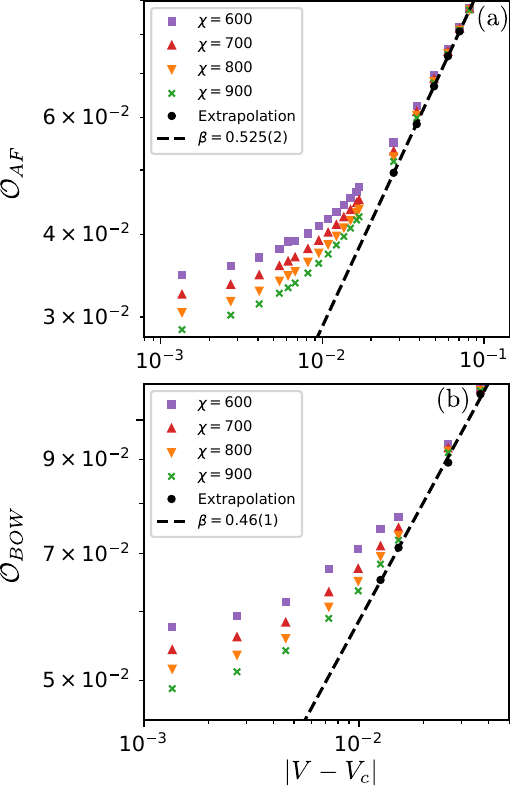}
    \caption{Scaling of the order parameters around the the AF-BOW transition point in appearing in Eq. \eqref{eq:ehm} for $U =4$, $J_z=t=1$ as a function of the bond dimension $\chi$. Both the (a) AF and the (b) BOW order parameters vanish continuously at transition. By extrapolating the limit $\chi\rightarrow\infty$, it is possible to extract the critical exponent $\beta_{AF},\beta_{BOW}\neq 1/8$, excluding an Ising transition.}
    \label{fig:AF-BOW_ord}
\end{figure}
On the other, $K_c$ shows a $L$-dependence compatible with a vanishing thermodynamic value for all the considered values of $V$, except at the transition point where the size dependence is absent. This strongly points in the direction that a continuous phase transition between the two LO regimes persists not only in the low energy limit. In order to enforce this aspect, we extract the correlation length $\xi_\nu(\chi)$ relative to the $\nu$ degrees of freedom~\footnote{The correlation length is extracted through the relation $\xi_\nu = -N/\log(\lambda_2)$ where $N$ is the number of sites of a unit cell and $\lambda_2$ is second highest eigenvalue of the transfer matrix.}. A genuine second-order phase transition can indeed be identified by considering the scaling of the correlation length $\xi_\nu$ as a function of the bond dimension $\chi$. In particular, one has to expect that in presence of a single point gap closing, the correlation length would show a $\chi$-dependence in the form of a cusp uniquely at that point. Fig.~\ref{fig:CDW_BOW}(b) shows this behavior for $\xi_c$, thus strongly supporting the continuous nature of this transition. Notably, this charge gap closing clearly supports the phenomenon of symmetry enriching. Indeed, based on known symmetry properties of SG model (see Appendix), the presence of a gapless sector is associated to a $U(1)\times U(1)$ (possibly $SU(2)$)  symmetry  in the same sector. It is straightforward to realize that such emergent symmetry relative to the charge degrees of freedom --not present either in BOW and CDW phases-- takes place at the CDW-BOW transition point.  
Further relevant information can be derived by calculating the decay of the correlators in $C_{CDW}(r)$ and $C_{BOW}(r)$. Fig.~\ref{fig:CDW_BOW}(c) clearly shows their expected algebraic decay at the transition point.  In addition, the same figure confirms that the spin parity parity correlator has long-range order, i.e. $C_P^s(r)\neq 0$ for $r\rightarrow\infty$.\\
Based on this analysis, we have strong indications that peculiar DQCPs can occur in one-dimensional interacting fermionic systems. As underlined, previous examples found indeed fully gapless DQCPs. As a consequence of the phenomenon of spin-charge separation, our analysis shows instead strong evidences on the presence of DQCPs with a finite gap and long-range order of a nonlocal order parameter.\\ 
In order to rule out the possibility to ascribe this phase transition in Ising universality class, in Fig.~\ref{fig:CDW_BOW_order} we report an analysis of the behavior of the local order parameters $\mathcal{O}_{CDW}$ and $\mathcal{O}_{BOW}$. Specifically, we are interested in extracting the critical exponent $\beta_{BOW/CDW}$ governing how the local BOW and CDW order parameters vanish at the transition point, i.e.
\begin{equation}
    \mathcal{O}_{BOW/CDW}(|V-V_C)\propto |V-V_C|^{\beta_{BOW/CDW}}.
\end{equation}
where $V_C$ is the critical value of the intersite interaction where we detect the charge gap closing. Crucially, while an Ising phase transition would formally imply $\beta_{BOW/CDW}=1/8$, our results based on an extrapolation in the limit  
$\chi\rightarrow\infty$, see \ref{app: extrap}, find different values of such critical exponent thus proving this phase transition as an example of a 1D DQCP. As pointed out, with respect to previously investigated DQCPs here such beyond Landau's transition points are further associated to a gapped excitation spectrum giving rise to long-range order of the spin parity operator.\\Notably, in agreement with previous finite-size studies on the same model ~\cite{Ejima2007}, we find the BOW-CDW phase transition becomes discontinuous for larger values of the interactions and therefore the partially gapped DQCPs disappear. However, this result is certainly not surprising. When increasing the contact interaction $U$, the mean-field treatment of the transition associated to negligible quantum fluctuations becomes indeed more accurate and a genuine first-order transition is expected. We prove this point for $U=8t$. In this case, the order parameters $\mathcal{O}_{BOW}$ and $\mathcal{O}_{CDW}$ exhibit a sharp jump at the critical point independent of the bond dimension $\chi$, see Fig.~\ref{fig:CDW_BOW_firstorder}(a), 
thus discarding the possibility of a single point gap closing. In addition, the Luttinger parameters $K_\nu$ reported in Fig.~\ref{fig:CDW_BOW_firstorder}(b) show both a clear discontinuity a the transition point and very small size dependent values thus compatible with a thermodynamic vanishing behaviour.  
\subsubsection{AF-BOW phase transition}\label{afbow}
\begin{figure}
    \centering
    \includegraphics[width=0.7\linewidth]{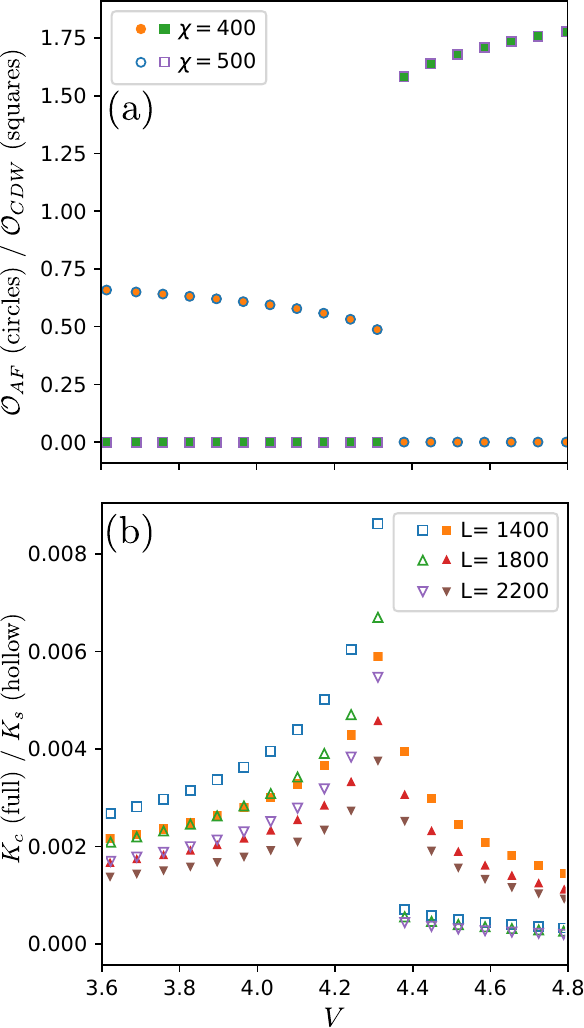}
    \caption{First order phase transition between the AF and the CDW phases appearing in Eq. \eqref{eq:ehm} for $U=8$ and $J_z=t=1$. (a) Values of the order parameters $\mathcal{O}_{BOW}$ and $\mathcal{O}_{AF}$ for different dimension $\chi$; (a) Values of the Luttinger parameter $K_c$ and $K_s$ at fixed bond dimension. The size $L$ refers to a finite MPS constructed out of $L/2$ unit cells.}
    \label{fig:AF-CDW}
\end{figure}
We now move to the analysis of the model in Eq.~\eqref{eq:ehm} for $J_z>0$ which phase diagram is shown in Fig. \eqref{phasediag}(b). As specified, a finite antiferromagnetic coupling allows for the breaking of the discrete spin flip symmetry. A part from slightly moving the transition points, we find indeed that the main role played by small $J_z$ is to replace in the Hamiltonian ground state the disordered MI with a locally ordered AF phase. Thanks to this aspect, we can now perform an analysis analogous to the one made in the previous section to investigate whether the AF-BOW transition point can also be classified as a partially gapped DQCP. Here, we focus on intermediate interactions $U/t=4$ and $J_z/t=1$ while varying $V/t$ and we again start by exploring the behavior of the Luttinger parameters. Our results in Fig.~\ref{fig:AF-BOW_crit}(a) find that at the transition point $K_c\approx 0$ \footnote{the increasing of $K_c$ for larger values of $V/t$ is due to the fact that the system is approaching the BOW-CDW transition} while $K_s$ shows a finite value not affected by different $L$. This analysis combined with the clear cusp of $\xi_s(\chi)$ shown in Fig.~\ref{fig:AF-BOW_crit}(b) demonstrates that the AF-BOW is also a continuous phase transition where the spin gap closes and $\Delta_c$ remains finite. Based of the already pointed out symmetry arguments, this results proves the emergence at the critical point of a $U(1)\times U(1)$ symmetry in the spin channel not present in either the BOW or the AF phase. 
Fig.~\ref{fig:AF-BOW_crit}(c) also demonstrates that such symmetry enriched transition point is associated to long-range order of the $C_P^c$ while the correlators $C_{AF}$ and $C_{BOW}$ decay algebraically. Thanks to these findings, we can again infer that the phase transition between the two locally ordered AF and BOW phases is continuous and thus supporting the possible appearance of a partly gapped DQCP. 
The results in Fig.~\ref{fig:AF-BOW_ord} confirm this intuition. As before, we can indeed show that local order parameters $\mathcal{O}_{AF}$ and $\mathcal{O}_{BOW}$ vanish continuously at the transition point. Moreover, through a $\chi$-extrapolation we calculate their critical exponent
\begin{equation}
    {\cal O}_{BOW/AF}(|V-V_C)\propto |V-V_C|^{\beta_{BOW/AF}},
\end{equation}
and we find $\beta_{AF}\sim0.525$ and $\beta_{BOW}\sim0.46$, again discarding the possibility of an Ising transition. As a consequence, this result proves again that the decoupling between spin and charge degrees of freedom can support the appearance a partly gapped DQCP which in the specific BOW-AF case is characterized by $\Delta_c\neq 0$ while having long-range order of $C_P^c(r)$.\\
Because of the specific values of the considered $J_z$, we do not find conclusive evidences of the possible disappearance of these DQCPs for larger interaction where, in analogy with previous case, the BOW-AF transition might become discontinuous. This is because the BOW phase disappears when increasing $U$ before the onset of a first-order BOW-CDW transition could happen. Nevertheless, it cannot be excluded that such a change of behaviour could happen for a different value of $J_z$.
\subsubsection{AF-CDW phase transition}\label{afcdw}
Finally, we study the AF-CDW phase transition. As already pointed out at the level of field theory, in order to find such transition to be continuous, both gaps have to vanish at the transition point. This aspect clearly rules out the possible presence of a partly gapped DQCPs while, in principle, it still allows for standard fully gapless DQCPs where all the correlators decay algebraically. As already discussed, the presence of such a DQCPs would require a fine tuning of the microscopic parameters in order to have a simultaneous closing of both the spin and the charge gap. 
In this regard, our numerical analysis finds that the AF-CDW transition is always first order for all values of $U/t$. In Fig.~\ref{fig:AF-CDW} we prove this for the case of $U/t=8$. As can be seen, we find indeed that both the order parameters ${\cal O}_{CDW}$ ${\cal O}_{AF}$ and the Luttinger parameters display a discontinuity at the transition point, see Fig.~\ref{fig:AF-CDW}(a) and (b) respectively. Moreover, while the former quantities are not affected by different $\chi$ values, the latter always show size dependence associated to almost vanishing values.
\section{Conclusions}
We unveiled a new type of deconfined quantum critical points. Contrary to previously known examples where they are associated to fully gapless excitations, we have shown that the phenomenon of spin and charge separation can generate partly gapped deconfined quantum critical points. Specifically, by means of analytical and numerical treatments, we have been able to prove that specific phase transitions connecting locally ordered fermionic phases are not captured by the Landau-Ginzburg-Wilson symmetry-breaking paradigm. Notably, at these phase transitions we find that only one gap closes while the other remains finite. Therefore, long-range order of specific nonlocal correlation functions occurs. In addition, we calculated the critical exponents governing how the local order parameters vanish. In such a way, we demonstrated that these transitions are one-dimensional realizations of deconfined quantum critical points with the intriguing feature of being only partly gapped. In conclusion, we also point out that at least one of these phase transitions, the one between the bond-order-wave and charge-density-wave, has been predicted to characterize the phase diagram of ultracold dipolar atoms in optical lattice \cite{DiDio2014,julia-farre2022}. Here, the presence of partially gapped deconfined quantum critical points can be probed by measuring the defect production \cite{Keesling2019,Xi_2022,Shu2022} predicted by the Kibble-Zurek mechanism \cite{Kibble1976,KIBBLE1980,Zurek1985,ZUREK1996,Polkovnikov2011,delCampo2014} in continuous transitions. In addition, quantum gas microscopy \cite{Gross2021} and noise correlators measurements \cite{gallegolizarribar2024} allow for an accurate detection of local density \cite{Endres2011} and spin \cite{Parsons2016} orderings and therefore for an accurate probing of nonlocal order parameters and locally ordered phases. As consequence, we believe that our results not only unveil a novel interesting type of deconfined quantum critical points but also pave the way towards their experimental detection, which represents a fundamental challenge solid state platforms. ~\cite{Zayed2017,Guo2020,Tao2022,Cui2023}.

\begin{acknowledgments}
We acknowledge useful discussions with M. P. A. Fisher, M. Oshikawa, G. Palumbo and L. Santos. ICFO group acknowledges support from: ERC AdG NOQIA; MCIN/AEI (PGC2018-0910.13039/501100011033, CEX2019-000910-S/10.13039/501100011033, Plan National FIDEUA PID2019-106901GB-I00, Plan National STAMEENA PID2022-139099NB-I00 project funded by MCIN/AEI/10.13039/501100011033 and by the “European Union NextGenerationEU/PRTR" (PRTR-C17.I1), FPI); QUANTERA MAQS PCI2019-111828-2); QUANTERA DYNAMITE PCI2022-132919 (QuantERA II Programme co-funded by European Union’s Horizon 2020 program under Grant Agreement No 101017733), Ministry of Economic Affairs and Digital Transformation of the Spanish Government through the QUANTUM ENIA project call – Quantum Spain project, and by the European Union through the Recovery, Transformation, and Resilience Plan – NextGenerationEU within the framework of the Digital Spain 2026 Agenda; Fundació Cellex; Fundació Mir-Puig; Generalitat de Catalunya (European Social Fund FEDER and CERCA program, AGAUR Grant No. 2021 SGR 01452, QuantumCAT \ U16-011424, co-funded by ERDF Operational Program of Catalonia 2014-2020); Barcelona Supercomputing Center MareNostrum (FI-2023-1-0013); EU Quantum Flagship (PASQuanS2.1, 101113690); EU Horizon 2020 FET-OPEN OPTOlogic (Grant No 899794); EU Horizon Europe Program (Grant Agreement 101080086 — NeQST), ICFO Internal “QuantumGaudi” project; European Union’s Horizon 2020 program under the Marie Sklodowska-Curie grant agreement No 847648; “La Caixa” Junior Leaders fellowships, La Caixa” Foundation (ID 100010434): CF/BQ/PR23/11980043. A. M. acknowledges financial support from the ICSC – Centro Nazionale di Ricerca in High Performance Computing, Big Data and Quantum Computing, funded by European Union – NextGenerationEU (Grant number CN00000013). M.R. acknowledges support from the Deutsche Forschungsgemeinschaft (DFG) via project Grant No. 277101999 within the CRC network TR 183 and the EU Quantum Flagship (PASQuanS2.1, 101113690). L. B. acknowledges financial support within the DiQut Grant No. 2022523NA7 funded by European Union – Next Generation EU, PRIN 2022 program (D.D. 104 - 02/02/2022 Ministero dell’Università e della Ricerca). Views and opinions expressed are, however, those of the author(s) only and do not necessarily reflect those of the European Union, European Commission, European Climate, Infrastructure and Environment Executive Agency (CINEA), or any other granting authority. Neither the European Union nor any granting authority can be held responsible for them.
\end{acknowledgments}

\appendix
\section{Field theory approach} \label{app:bosonization}

The standard field theory treatment of 1D lattice systems of correlated fermions is based on a weak coupling approach, known as bosonization. As this technique has been largely employed, see ~\cite{Gogolin1998,Giamarchi2004} for exhaustive references, here we just summarize the relevant steps needed to derive the low energy properties of a many-body fermionic system with short range interaction within Abelian bosonization. The starting point consists in defining the continuum limit by replacing the discrete sum over sites $j$ with integrals over the coordinate $x = ja$, i.e. $\sum_j\rightarrow \frac{1}{a} \int dx$ with $a$ lattice spacing. Thanks to the peculiar nature of the one dimensional Fermi surface consisting of just two disconnected points $\pm k_F$, usual fermionic operators $c_{j\sigma}, c_{j\sigma}^\dagger$, with $\sigma=\uparrow,\downarrow$ labeling the spin orientation, can be rewritten in terms of right ($R$) and left ($L$) fermionic fields $\psi_{\chi\sigma}(x)$ ($\chi=R,L$), 
\begin{equation}\label{def}
c_{j\sigma}\rightarrow\sqrt{a} \left [e^{ik_F x} \psi_{R\sigma} (x)+ e^{-ik_F x} \psi_{L\sigma} (x)\right ] \quad .
\end{equation}
The bosonization procedure amounts to rewriting the above fields as appropriate exponentials of bosonic fields $\phi_{\chi\sigma}(x)$, multiplied by a Klein factor $\eta_{\chi\sigma}$ to reproduce their correct anticommutation algebra: $\psi_{\chi\sigma}(x)=\frac{\eta_{\chi\sigma}}{\sqrt{2\pi\alpha}}e^{i \sqrt{4\pi}\chi\phi_{\chi\sigma}(x)}$. It is then customary to introduce first the linear combinations $\phi_\sigma(x)=\phi_{R\sigma}(x)+\phi_{L\sigma}(x)$ and $\theta_\sigma(x)=\phi_{R\sigma}(x)-\phi_{L\sigma}(x)$; and finally their superpositions $\varphi_c(x)=\phi_\uparrow+\phi_\downarrow$, and $\varphi_s(x)=\phi_\uparrow-\phi_\downarrow$, and similarly $\vartheta_c = \theta_{\uparrow} + \theta_{\downarrow}$ and $\vartheta_s = \theta_{\uparrow} - \theta_{\downarrow}$ . At half filling $k_F=\pi/(2 a)$ the resulting mapping reads:
\begin{equation}
\psi_{\chi\sigma}=\frac{\eta_{\chi\sigma}}{\sqrt{2\pi\alpha}} e^{i \sqrt{\frac{\pi}{2}}\left [\chi \varphi_c(x)+\vartheta_c(x)+\sigma \left (\chi\varphi_s(x)+\vartheta_s(x)\right )\right ]} \quad ,
\label{field}
\end{equation}
where, with an abuse of notation, $\chi$ and $\sigma$ in the exponent now stand for signs: $R=+$ and $L=-$, $\uparrow=+$ and $\downarrow=-$. Moreover $\alpha\sim a$ is an ultraviolet cutoff, and appropriate commutation relations hold between the bosonic fields. The procedure outlined above allows describing the low energy behavior of most one dimensional fermionic systems. In case of conserved total magnetization and particle number a fermionic system is described by two decoupled sine-Gordon models  ${\cal H}=\sum_{\nu=c,s}{\cal H}_\nu^{SG}$ in the spin ($s$) and charge ($c$) degrees of freedom. Explicitly: 
\begin{equation}
\begin{split}
 {\cal H}^{SG}_\nu = \frac{1}{2} \int dx   \bigg[ v_\nu K_\nu (\nabla \vartheta_\nu(x))^2  +\frac{v_\nu}{K_\nu}  (\nabla \varphi_\nu(x))^2\\
 +\frac{g_\nu}{\pi^2 a^2} \cos(\sqrt{8\pi} \varphi_\nu(x))\bigg]
 \quad .
 \end{split}
 \label{SG_app}
 \end{equation}
Notice that here $v_\nu$, $K_\nu$ and $g_\nu$ are the excitation velocities, Luttinger parameters, and coupling amplitudes respectively, which depend on the microscopic Hamiltonian parameters. Other cosine terms, which may appear within each sine-Gordon Hamiltonian ${\cal H}_\nu^{SG}$ or as coupling terms between the charge and spin sectors, are typically neglected in the low-energy theory. This is because they involve faster oscillations in the bosonic fields and, accordingly, acquire larger scaling dimensions. From a renormalization group (RG) perspective, such terms are therefore irrelevant and do not affect the universal low-energy physics.

\subsection{Derivation of gapped phases and symmetry breaking}

In Eq.~\eqref{SG_app}, the competition between the quadratic and the cosine terms determines the properties of the system. Specifically, in the first line fluctuations of the fields are promoted, whereas the cosine term in the second line favors the pinning of the field $\varphi_\nu(x)$ to a $x$ independent value minimizing the system energy. The resulting phase diagram can be established by means of a well known RG analysis ~\cite{Gogolin1998} of the model in Eq.~\eqref{SG_app}. 
In each channel the cosine term turns out to be irrelevant if the coupling $g_\nu$ satisfies  
\begin{equation}
2 \pi v_\nu (K_\nu-1)\geq |g_\nu| \quad .
\label{RG}
\end{equation}
When the cosine term becomes relevant instead, $\varphi_\nu(x)$ prefers to pin to an appropriate constant value, and a gap $\Delta_\nu$ opens in the corresponding $\nu$-channel. Since the gapless excitations are suppressed, the long wavelength fluctuations and the associated power law behavior of correlations captured by $K_\nu$ are also suppressed: $K_\nu$ effectively flows to $0$ in the $\nu$-channel. 
\\In general, depending on the negative or positive sign of $g_\nu$ in Eq.~\eqref{RG}, the pinning values of $\varphi_\nu$ are respectively $0$ and $\sqrt{\pi/8}$. In presence of total charge/spin vector conservation  of the microscopic lattice Hamiltonian though, the RG flow in the corresponding channel is restricted to solely the line $2 \pi v_\nu (K_\nu-1)=g_\nu$, implying that the cosine term can become relevant only for $g_\nu<0$ ($\varphi_\nu=0$). In this case $K_\nu=1+\frac{g_\nu}{2 \pi v_\nu}$: the theory is gapless for $K_\nu > 1$ and gapped for $K_\nu < 1$. More generally, rooted in the total charge/spin conservation of the lattice Hamiltonian is the global $U(1)$ symmetry
\begin{equation}
\vartheta_\nu\rightarrow \vartheta_\nu+const \quad
\end{equation}   
and a further solution to Eq.~\eqref{RG} can be found, corresponding to the choice $\varphi_\nu=\sqrt{\pi/8}$ ~\footnote{Indeed, in this more general case the Luttinger parameters can be written in the form $K_\nu= \frac{1}{2 \pi v_\nu}(1+g_\nu-\kappa_\nu)$, allowing a further solution  for $\kappa_\nu>0$}.
\\When the cosine term is irrelevant, the effective Hamiltonian reduces to a Gaussian model, exhibiting the further $U(1)$ symmetry
\begin{equation}
\varphi_\nu(x) \rightarrow \varphi_\nu(x)+const \quad ,
\end{equation}
so that the full symmetry in a gapless channel will be enriched to $U(1)\times U(1)\cong SU(2)$ ~\footnote{notice that the effective $SU(2)$ symmetry holds in the case of $SU(2)$ symmetric choices of the microscopic parameters of a Hamiltonian}. When the cosine term becomes relevant though, the above symmetry becomes discrete, namely
\begin{equation}
\varphi_\nu(x)\rightarrow \varphi_\nu (x)+\sqrt{\frac{\pi}{2}} \quad . \label{Z2}
\end{equation}
Relevantly, also other discrete like the translational, spin-flip and inversion symmetries characterize ${\cal H}$, see for instance ~\cite{Nakamura2000,Mudry2019}. According to the Landau's picture ~\cite{Landau_ssb}, some of these could be spontaneously broken generating LO phases captured by a corresponding local order parameter.

\subsection{Bosonization of the microscopic model}
The bosonization technique of the microscopic model
\begin{eqnarray}
H=&-&t\sum_{j,\sigma}(c_{j,\sigma}^\dagger c_{j+1,\sigma}+h.c.)+U\sum_jn_{j,\uparrow}n_{j,\downarrow}\nonumber\\
&+&\sum_{j}(VS_j^cS_{j+1}^c+J_zS_j^sS_{j+1}^s)
\end{eqnarray}
provides the following quantities
\begin{equation}
    \begin{aligned}
    v_\nu &= 2 a t (2-K_\nu),\cr
    K_c&=  1-\frac{1}{4\pi t} \bigl (U+6V+2J_z\bigr), \cr
    K_s&=  1-\frac{1}{4\pi t} \bigl (-U+2V+6 J_z\bigr), \cr
    g_c&= -\bigl(U-2V+2J_z\bigr ) a  = -g_s.
    \end{aligned}
    \label{paraLL}
\end{equation} 
Based on the previous results, it is thus possible to predict that the transitions between LO phases occur independently in the two channels for $g_\nu=0$. In this case, since $g_c=-g_s$, they identify the same line, namely  
\begin{equation}
U+2J_z=2V \quad . \label{Jztrans_app}
\end{equation} 

\section{Extrapolation of order parameters around criticality} \label{app: extrap}
\begin{figure}
    \centering
    \includegraphics[width=0.8\linewidth]{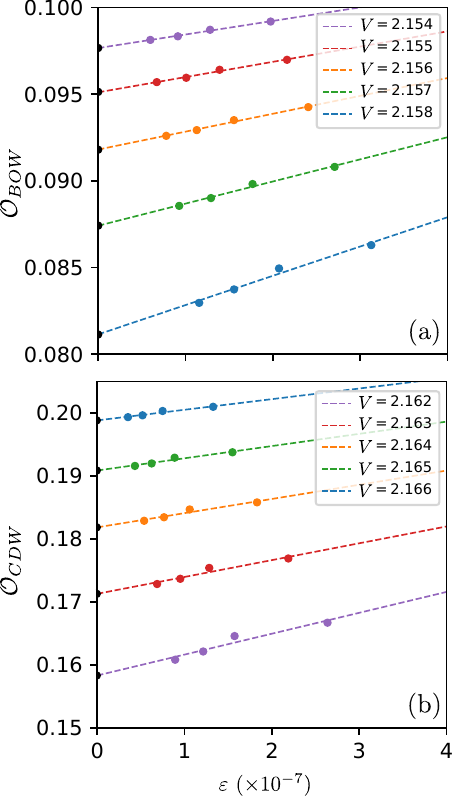}
    \caption{Extrapolation of the order parameters (a) $\mathcal{O}_{BOW}$ and (b) $\mathcal{O}_{CDW}$, for $\chi\rightarrow\infty$, as a function of the density interaction $V$. We use results computed at $\chi=550,650,750,850$. The extracted values (black dots) correspond to the extrapolated points in in Fig.\ref{fig:CDW_BOW_order}.}
    \label{fig:extrap}
\end{figure}
To address the behavior of the order parameters around the DQCPs in Sections \ref{bowcdw} and \ref{afbow}, we have to properly extrapolate their behavior in the limit of infinite bond dimension $\chi \rightarrow \infty$. This is crucial as the scaling of these quantities around the critical point leads us to obtain critical exponents incompatible with a transition in the Ising universality class. This, in turn, corroborates our conclusion that we indeed find a DQCP. However, the convergence criterion usually used in VUMPS, only assures that we find a variational minimum in the manifold of the finite bond dimension ansatz in consideration \cite{Zauner2018}. This does not readily give a measure of the quality of the approximation when compared to the physical ground state. 
\\To overcome this problem we proceed as in standard DMRG, by performing an extrapolation as a function of the discarded Schmidt values after performing a truncation during the optimization process \cite{schollwock_density-matrix_2011}. 
\\
%
In particular, to obtain this truncation error, it is sufficient to apply the Hamiltonian $H$ projected on the subspace of the tensors currently being optimized, obtaining a state of bond dimension $d\chi$, where $d=4$ in the case of spinful fermions. We can then obtain the new ground state in this subspace by performing an exact diagonalization. Finally, by truncating this state back to bond dimension $\chi$ we can obtain the error. This procedure is straightforward in the case of two-site DMRG, where the growing of the bond dimension is dynamically performed during the optimization sweep. In the case of VUMPS, however, the bond dimension is kept fixed during the optimization process. We then proceed in the following way: (i) we perform a subspace expansion to express the converged ground state at bond dimension $\chi$ using a larger bond dimension $4\chi$ \cite{hubig2015}. (ii) We perform one sweep of the VUMPS optimization, equivalent to a single application of $H$ to the unit cell and an exact diagonalization in this subspace, (iii) we manually compute the error 
\begin{equation}
    \varepsilon = \sum_{i=\chi}^{4\chi} S^2_i,
\end{equation}
where $S_i$ are the Schmidt values along the bond on which the Hamiltonian is applied. This procedure follows the one explained in Appendix A4 of \cite{Zauner2018}.
\\In Fig.\ref{fig:extrap} we show, as an example, the value of the order parameters $\mathcal{O}_{BOW/CDW}$ as a function of $\varepsilon$ for the BOW-CDW transition showed in Sec. XX. From the extrapolated values of $O_{BOW/CDW}$ we then extracted the critical exponents $\beta_{BOW/CDW}$ by doing a linear fit of the function 
\begin{equation}
\begin{split}
    \log_{10}\left(\mathcal{O}_{BOW/CDW}(V-V_c)\right) = \\= \beta_{BOW/CDW}\log_{10}\left(|V-V_c|\right) + \alpha
    \end{split}
\end{equation}
using the least squares method. Here, $\alpha$ is a non-universal parameter, not relevant for our analysis. In this way we identified the value of $\beta$ with an accuracy of the third decimal digit, excluding a value $\beta=1/8$. We used the same process for the AF-BOW transition to extract the limiting values at criticality of $\mathcal{O}_{AF}$ and $\mathcal{O}_{BOW}$.

\bibliography{main_paper.bib}

\end{document}